# Transiting Exoplanets from Sharjah Astronomical Observatory (SAO-M47): The Exoplanet HAT-P-25 b Using L & V Filters

Mohammad F. Talafha[1(✉)], Mashhoor A. Al-Wardat[1,2], Ammar E. M. Abdulla[1,2], and Hamid M. Al-Naimiy[1,2]

[1] Sharjah Academy of Astronomy Space Science and Technology, University of Sharjah, 27272 Sharjah, United Arab Emirates
`mtalafha@shrjah.ac.ae`
[2] Physics Department, Science College, University of Sharjah, 27272 Sharjah, United Arab Emirates

**Abstract.** In this study, we conduct a comparative analysis of observations carried out on the exoplanet HAT-P-25b at the Sharjah Astronomical Observatory (SAO). We have employed two distinct filters, namely, the Luminoso ($L$) and Visual ($V$) filters. Our research conducted aims to discern any variations in transit depth or exoplanet size resulting from the use of these different filters.

The primary focus of this study is to determine the exoplanet's size relative to its host star using the transit method. The application of different filters was expected to introduce subtle variations in size, influenced by factors such as the exoplanet's atmosphere. Notably, our findings reveal that the exoplanet's size appears larger when observed through the L filter compared to the V filter.

Throughout the analytical process, we employed the TRASCA model to determine the transit depth for each epoch. Fixed parameters, including the orbital period of the exoplanet ($P$, measured in days) and the transit duration (measured in minutes), were utilized in these calculations. Our results indicate that the transit depths observed with the $L$ filter were greater than those with the $V$ filter, measuring 0.0238 magnitudes and 0.0200 magnitudes, respectively. These values deviate from the reference result of 0.0204 magnitudes.

## 1 Introduction

Over the last two decades, the identification of exoplanets, facilitated by the transit method, has swiftly transformed our comprehension of the cosmos, more than 60% of the discovered extrasolar planets were identified using this method (1) Moreover, there has been a rapid growth in the number of exoplanet candidates, a surge which was facilitated through surveys such as the TESS & Kipler satellite. The initial discovery of HAT-P-25b was made by Quinn (2012). Subsequent refinements to the physical and orbital parameters of the system were conducted by Wang (2018). (2) No photometric or spectroscopic studies have been conducted on the HAT-P-25 star-planet system other than the discovery paper and the paper by Wang et al. (2018). (3) In this study, we introduce new photometric follow-up observations using different filters. The parameters of the exoplanet, as listed in the discovery publication, report that it is a transiting extrasolar planet orbiting the $V = 13.19$, $G5$ dwarf star $GSC1788 - 01237$, with a period $P = 3.652836 \pm 0.000019 d$, and transit duration $0.1174 \pm 0.0017 d$. The host star has a mass of $1.01 \pm 0.03 M_\odot$, and radius of $0.96 \pm^{0.05}_{0.04}$ $R_\odot$, an effective temperature of $5500 \pm 80$ K, and metallicity [Fe/H] = $+0.31 \pm 0.08$. The planetary companion has a mass of $0.567 \pm 0.022 M_J$, and radius of $1.190 \pm^{0.081}_{0.056}$ $R_J$, yielding a mean density of $0.42 \pm 0.07$ $gcm^{-3}$ (4). Also, the corrected transit depth found was $(0.01626 \pm 0.00035)$ using Sloan i band filter (2), observed by the Fred Lawrence Whipple Observatory (FLWO), as shown in Fig. 1. The TESS telescope has observed this target multiple times, as shown in Fig. 2. The multi-filter photometric observation was used to recognize the different responses for the transit depth. Observations in this spectral band are crucial for examining the presence of high hazes in the atmosphere. Rayleigh scattering gains significance at shorter wavelengths, potentially resulting in larger observed radii within this band. (5).

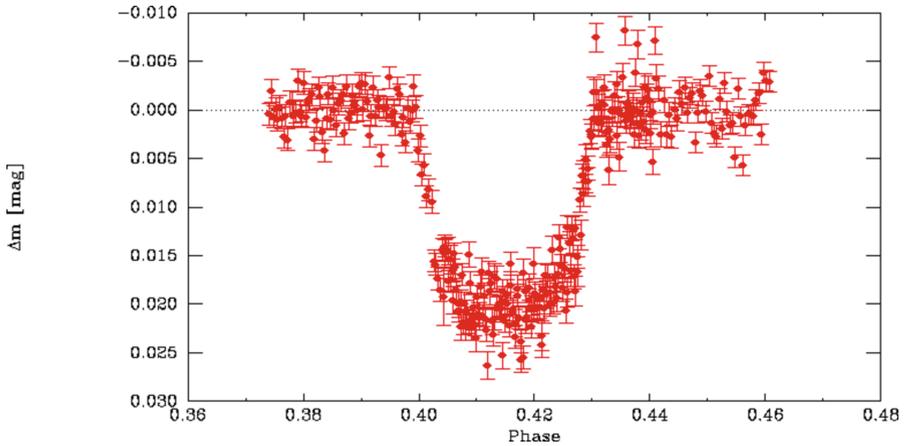

**Fig. 1.** The transit light curve for HAT-P-25b by Sloan i band filter (650 nm–85 m), using the Fred Lawrence Whipple Observatory (FLWO)

The resulting light curve is shown in Figs. 3 and 4. Although there is more dispersion in the data taken with the $V$ filter compared to the $L$ filter data, a flux depth of approximately 2% or more is observed. The best-fit of $Rp/R*$ values obtained are listed in Table 2.

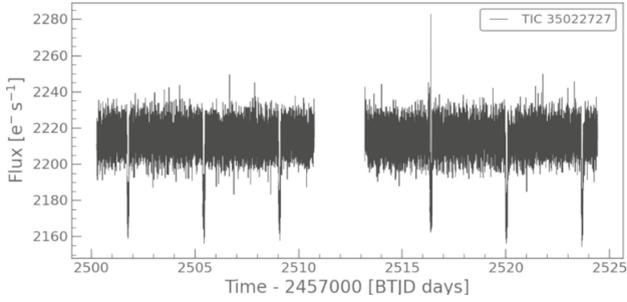

**Fig. 2.** Multi-epoch light curves by TESS space telescope observing HAT-P-25b found in sector 44, released in 2021.

## 2  Observations and Data Reduction

The observations were conducted at the Sharjah Astronomical Observatory (SAO-M47) employing a 431 mm aperture telescope equipped with two distinct optical filters: a Luminous filter *L* spanning the spectral range from 420 nm to 685 nm that provides 98% transmission over the entire visible spectrum from 420 to 685 nm, and a *V* filter covering the range of 500 nm to 700 nm. The *V* filter demonstrates a level of transmittance centered around its peak wavelength of about 530 nm. However, it allows only around 63% of the incident light at that specific wavelength to traverse through the filter. The SAO's optical system incorporates a CDK telescope configuration along with an *SBIGSTX − 16803CCD* Camera $4K \times 4K$, delivering a field of view measuring $43\prime \times 43\prime$ and an angular pixel scale of 0.63 arcseconds per pixel.

Standard data processing procedures were applied to the compiled observations, involving dark frame subtraction, bias correction, and flat-field calibration. Aperture photometry was subsequently performed utilizing *MaximDL* software. The transit light curves were derived using differential photometry techniques (6).

The fundamental parameters of the exoplanet were determined based on the TRASCA (7) model, assuming a fixed radius for the host star denoted as $R\odot$ (solar radii), a fixed period for the exoplanet $P$ (in days), and the transit duration (in minutes).

The observation commenced by consulting the Exoplanet Transit Database to ascertain the timing of observations and the altitude of the exoplanet. The initial observation, conducted with the Luminous filter, took place on November 8, 2021 (Julian Heliocentric Date: JHD 2459527.29170), featuring an exposure duration of 150 s during which a total of 73 individual images were acquired.

A year later, the second observation was performed utilizing the *V* filter on November 12, 2022 (Julian Heliocentric Date: JHD 2459896.22108), with a longer exposure time of 360 s. A total of 46 images were collected during this observation. All details of these observations are listed in Table 1 (Figs. 5 and 6).

**Table 1.** The observation parameters for each night of observation.

| Parameters | | | | | | Air mass | |
|---|---|---|---|---|---|---|---|
| Filter | Exposure sec | Binning | # Frames | 1/SNR | Binning | Start | End |
| L | 150 | 1 | 73 | 0.003 | $1 \times 1$ | 1.0658 | 1.3243 |
| V | 360 | 1 | 42 | 0.007 | $1 \times 1$ | 1.3082 | 1.0810 |

The analytical illustrations generated by the TRASCA Model depict consistency in outcomes regarding the planet's rotation period and transit duration when contrasted with other observations at the ETD site. Additionally, these illustrations show a clear disparity in the depth of the transit observed between the L and V filters.

## 3  Results

Utilizing the TRESCA model for data analysis and light curve generation, with fixed parameters including mid-transit and transit duration, we examined variations in transit depth. A comparative analysis of observations conducted with *L* and *V* filters revealed distinct differences in transit depth between them.

Moreover, transit depth (bottom) showed the transit depth of observations made by the L filter to be larger than ones by the *V* filter Fig. 3 and Fig. 4. On the other hand, the Top and middle plots show the remarkable agreement with reference results and correspondence in the calculated and observed results of observing the planet using the *V* filter and filter *L*, , and this is evident in all parameters (O-C, Duration time, and transit depth). Table 2 lists all the results of the observations that have been shared on the Exoplanet Transit Database website[1].

---

[1] http://var2.astro.cz/ETD/.

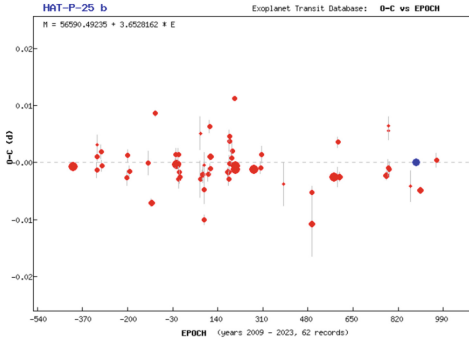
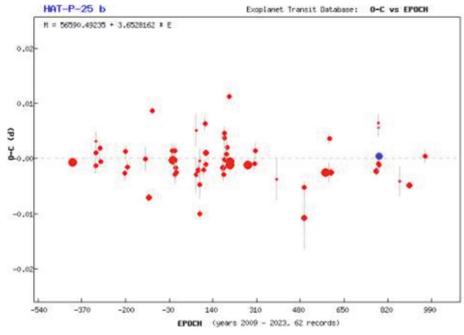
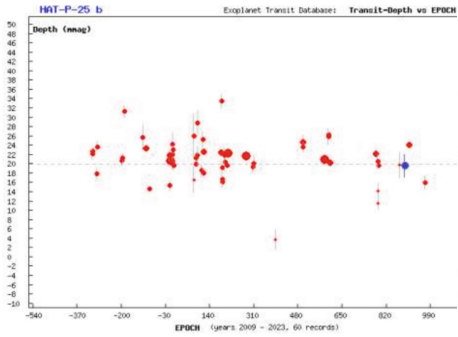
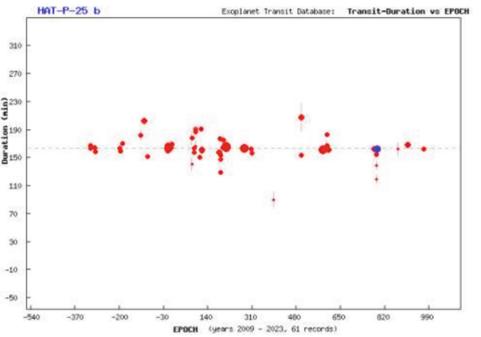
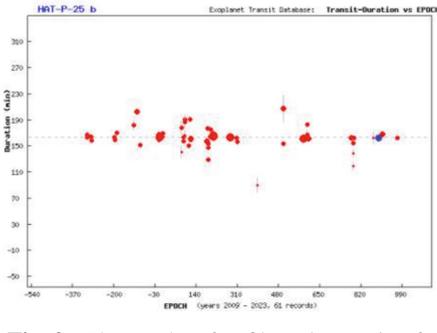
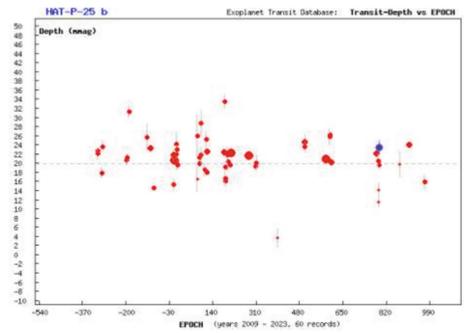

**Fig. 3.** The results of V filter observation for HAT-P-25b by TRESCA. Top: The fixed period of the exoplanet. Middle: Fixed result for the transit duration. Bottom: Depth of transit

**Fig. 4.** The results of L filter observations for HAT-P-25b. Top: The fixed period of the exoplanet. Middle: Fixed result for the transit duration. Bottom: Depth of transit.

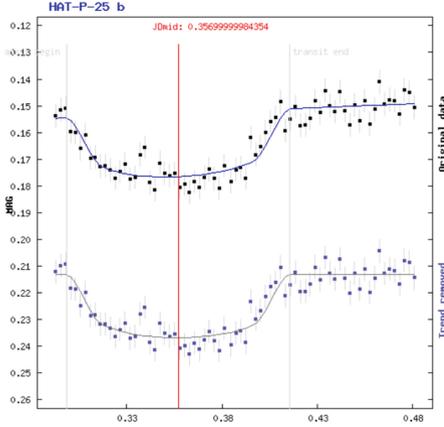
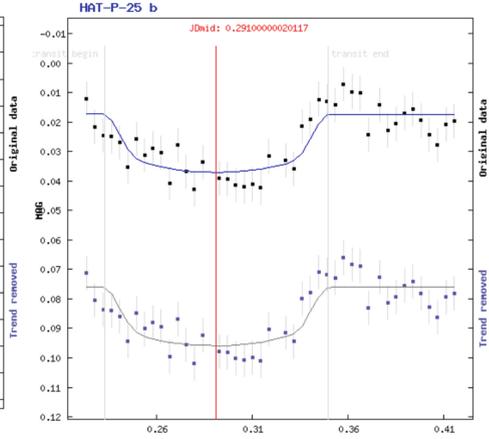

**Fig. 5.** Light curves in the L filter for HAT-P-25b.

**Fig. 6.** Light curve taken by V filter for HAT-P-25b.

**Table 2.** Results for HAT-P-25b observations by TRESCA model.

| Filter | Date | Duration (min) | Epoch | Depth | $R_p/R_*$ |
|---|---|---|---|---|---|
| Luminous (L) | 8/11/2021 | Fix | Fix | 0.0238 | 0.1542 |
| Visible (V) | 12/11/2022 | Fix | Fix | 0.0200 | 0.1414 |
| Reference (ETD) | | 169 min | --------- | 0.0204 | 0.1482 |

## 4   Conclusion

This study marks the inaugural presentation of a light curve of HAT-P-25b in the L filter. The significance of observations across various spectral bands lies in their ability to investigate the presence of high hazes in the atmosphere. This importance stems from the increased relevance of Rayleigh scattering at shorter wavelengths, potentially resulting in larger observed radii in the shortwave band (5). Transit duration and mid-transit were systematically fixed into the TRASCA Model as reference points to enable a rigorous comparison of transit depth outcomes between the two filters. The exciting results emphasize the importance of conducting more observations using various filters. These additional investigations are essential for detecting and quantifying the dynamic alterations in the depth of the light curve using different filters, and more photometric and transmission spectrum observations are needed.

## References

Neilson, H.R., et al.: Limb darkening and planetary transits: testing center-to-limb intensity variations and limb-darkening directly from model stellar atmospheres. Astrophys. J. **845**, 12 (2017)

Wang, X.Y., et al.: Transiting Exoplanet Monitoring Project (TEMP). IV. Refined system parameters, transit timing variations, and orbital stability of the transiting planetary system HAT-P-25. Public. Astron. Soc. Pac. **130**(988) (2018)

Erdem, A., Öztürk, O.: New photometric observations of HAT-P-25b and WASP-11b/HAT-P-10b. In: AIP Conference Proceedings (2018)

Quinn, et al.: Transiting Exoplanet Monitoring Project (TEMP). IV. Refined system parameters, transit timing variations and orbital stability of the transiting planetary system HAT-P-25. Astrophys. J. **745**, 9 (2012)

Ricci, D., et al.: Multifilter transit observations of WASP-39b and WASP-43b with Three San-Pedro Mártir telescopes. Publ. Astron. Soc. Pac. **127**(948), 143 (2015)

Henry, G.W.: Techniques for automated high-precision photometry of sun-like stars. Astron. Soc. Pac. **111**(761) (1999)

Stanislav, P., Luboš, B., Ondřej, P.: Exoplanet transit database. Reduction and processing of the photometric data of exoplanet transits. New Astron. **15**(3), 297–301 (2010)